\documentclass[aps,onecolumn]{revtex4-1}

\usepackage{amsmath}
\usepackage{amssymb}
\usepackage{graphicx}
\usepackage{bm}
\usepackage{epstopdf}
\usepackage{hyperref}
\hypersetup{
    colorlinks=true,
    linkcolor=blue,
    filecolor=magenta,      
    urlcolor=blue,
    citecolor=red,
}

\newcommand{\ra}{\rightarrow}
\newcommand{\longra}{\longrightarrow}
\newcommand{\arrA}{\overset{A}{\longra}}
\newcommand{\arrB}{\overset{B}{\longra}}
\newcommand{\U}{{\mathcal{U}}}
\renewcommand{\H}{{\mathcal{H}}}
\renewcommand{\L}{{\mathcal{L}}}

\begin{document}

\title[Bayesian Approach to  Naming Game]{A Bayesian Approach to the Naming Game Model} 
\vspace{1cm}
\author{\href{https://www.etis.ee/CV/Gionni_Marchetti/eng?lang=ENG}{\normalsize Gionni Marchetti}}
\author{\href{https://www.etis.ee/Portal/Persons/Display/05ca5332-db93-42e6-a373-a15f0ebf1693?tabId=CV_ENG&lang=ENG}{Marco Patriarca}}\thanks{{\normalsize\tt marco.patriarca@kbfi.ee}}
\author{\href{https://www.etis.ee/CV/Els_Heinsalu/eng?lang=ENG}{Els Heinsalu}}
\address{{\normalsize\it \href{https://kbfi.ee/?lang=en}{NICPB -- National Institute of Chemical Physics and Biophysics, Tallinn, Estonia}}}

\vspace{2cm}

\begin{abstract}
\textbf{Abstract}. 
{\normalsize 
We present a novel Bayesian approach to semiotic dynamics, which is a cognitive analogue of the naming game model restricted to two conventions.
The one-shot learning that characterizes the agent dynamics in the basic naming game is replaced by a word-learning process, in which agents learn a new word by generalizing from the evidence garnered through pairwise-interactions with other agents. 
The principle underlying the model is that agents --- like humans --- can learn from a few positive examples and that such a process is modeled in a Bayesian probabilistic framework. 
We show that the model presents some analogies but also crucial differences with respect to the dynamics of the basic two-convention naming game model.
The model introduced aims at providing a starting point for the construction of a general framework for studying the combined effects of cognitive and social dynamics.
}
\end{abstract}

\maketitle


\section{Introduction}

A basic question in complexity theory is how the interactions between the units of the system lead to the emergence of ordered states from initially disordered configurations \cite{Castellano-2009a,Baronchelli-2018a}.
This general question concerns phenomena ranging from phase transitions in condensed matter systems and self-organization in living matter to the appearance of norm conventions and cultural paradigms in social systems.
Various models were used in order to study social interactions and cooperation, e.g. models of condensed matter systems (such as spin systems), statistical mechanical models (e.g. based on the master equation), ecological competition models \cite{Castellano-2009a}, many-agents game-theoretical models \cite{Xia-2017a,Xia-2018a,Zhang-2017a}.
Opinion dynamics and cultural spreading models represent suitable theoretical frameworks for a quantitative description of the emergence of social consensus \cite{Baronchelli-2018a}.

In this respect, the emergence of human language remains a challenging,  multi-fold  question, related in turn to biological, ecological, social, logical, and cognitive aspects \cite{Mufwene-2001a,Lass1997,Berruto-2004a,Edelman-2007a,Tenenbaum-1999}.
Language dynamics \cite{Wichmann-2008b,Wichmann-2008c} has provided models describing phenomena of language competition and change that focus on the mutual interactions of linguistic traits (sounds, phonemes, grammatical rules, or languages understood as fixed entities) under the influence of ecological and social factors, modeling such interactions in analogy to biological competition and evolution.

However, the basic learning process of a word has a complex dynamics due to its cognitive dimension.
In fact, learning a word means to learn a \emph{concept} (understood as a pointer to a subset of objects, see Refs. \cite{Tenenbaum-1999, Tenenbaum-2000a,Xu-2007a}) and a linguistic label ---for example the \textit{name} of the object--- used for communicating the concept.
The double concept$\leftrightarrow$name nature of words has been studied through semiotic dynamics models, such as the models of Hurford \cite{Hurford-1989a} and Nowak \cite{Nowak-1999a} (see also \cite{Nowak-2000a,Trapa-2000a}) and the naming game (NG) model \cite{Baronchelli-2006c,guanrong2019}.

In the basic version of the model of Nowak \cite{Nowak-1999a}, the language spoken by each agent $i$ ($i = 1,\dots,N$) is defined by two personal matrices, representing the links of a bipartite network joining $Q$ names and $R$ concepts: 
(1) an active matrix $\U^{(i)}$ representing the concept$\,\ra\,$name links, where the element  $\U^{(i)}_{q,r}$ ($q \in (1,Q), ~ r \in (1,R)$) gives the probability that agent $i$ will utter the $q$th name to communicate the $r$th concept;
(2) the passive matrix $H^{(i)}$, representing the name$\,\ra\,$concept links, in which the element $\H^{(i)}_{q,r}$ represents the probability that an agent interprets the $q$th name as referring to the $r$th concept.
In the models of  Hurford and in the model of Nowak, the languages of each individual evolve with time according to a game-theoretical dynamics, with agents gaining a reproductive advantage if their matrices have a higher communication efficiency.
These studies have achieved interesting results, such as the emergence of non-ambiguous one-to-one links between objects and sounds, and explain why homonyms are more frequent than synonyms
\cite{Hurford-1989a,Nowak-1999a,Nowak-2000a,Trapa-2000a}.

In the NG model \cite{Baronchelli-2006c, guanrong2019} there is only one concept ($R=1$) that can be linked to a set of $Q > 1$ different names.
The model can be reformulated through the agents' lists $\L_i$ of the name$\leftrightarrow$concept connections known to each agent $i$.
In the case of two-conventions models, where the conventions are the names $A$ and $B$, the list of the $i$th agent can be $\L_i = \emptyset$ (no connection), $\L_i = (A)$ or $(B)$ (one name is known), or $\L_i = (A,B)$ (both name$\leftrightarrow$concept connections are known).

Extending semiotic dynamics models is not trivial and already two-opinion variants of the NG model, taking into account committed groups, show a remarkable phase diagram \cite{Xie-2012a}; and trying to describe actual cognitive effects requires entirely new features \cite{Fan-2018a}.
This paper presents a minimal model to study the interplay of the cognitive and social dynamical dimensions, assuming for simplicity the two-conventions NG model as a semiotic framework \cite{Baronchelli-2016a,guanrong2019} and making a cognitive generalization within the experimentally validated Bayesian framework of \cite{Tenenbaum-1999} (see also Refs. \cite{Tenenbaum2001, Tenenbaum-2000a,Griffiths2006, Xu-2007a, Perfors-2011a, Lake2015}).
In that framework, an individual can learn a concept from a small number of examples, a most remarkable feature of human learning \cite{Tenenbaum-1999,Tenenbaum-1999b,Tenenbaum-2011a}, to be contrasted with machine learning algorithms, which require a large amount of examples for generalizing successfully \cite{barber2012,Murphy-2012a,Evgeniou-2000a}. 

The paper is organized as follows. 
The new model is introduced in Sec. \ref{sec:gamerules}.
In Sec.~\ref{sec:results}, we present and discuss the features of the semiotic dynamics emerging from the numerical simulations and quantitatively compare them with those of the two-conventions NG model. 
Future directions in the study of the interplay of the cognitive and the social dynamics are outlined in Sec.~\ref{sec:conclusion}.


\section{A Bayesian learning approach to the naming game}
\label{sec:gamerules}

\subsection{The two-conventions naming game model}
\label{2cng}

Before introducing the new model, we recall the basic 2-conventions NG model \cite{Castello2009}, in which there is a single concept $C$, corresponding to an external object, and two possible names (synonyms) $A$ and $B$ for referring to $C$. 
Thus, the possibility of homonymy is excluded \cite{Baronchelli-2016a}. 
Each agent $i$ is equipped with the list $\L_i$ of the names known to the agent.
We assume that at $t = 0$ each agent $i$ knows either $A$ or $B$ and has therefore a list $\L_i = (A)$ or $\L_i = (B)$, respectively.

During a pair-wise interaction, an agent can act as a speaker, when conveying a word to another agent, or as a hearer, when receiving a word from a speaker.
One can think of an agent conveying a word as uttering a name, e.g. $A$, while pointing at an external object, corresponding to concept $C$: 
thus, the hearer records not only the name $A$ but also the name$\leftrightarrow$concept association between $A$ and $C$.
At a later time $t > 0$, the list $\L_i$ of the $i$th agent can contain one or both names, i.e., $\L_i = (A)$, $(B)$, or $(A,B)$.

\begin{figure}[h!]
\begin{center}
\includegraphics*[width=17cm]{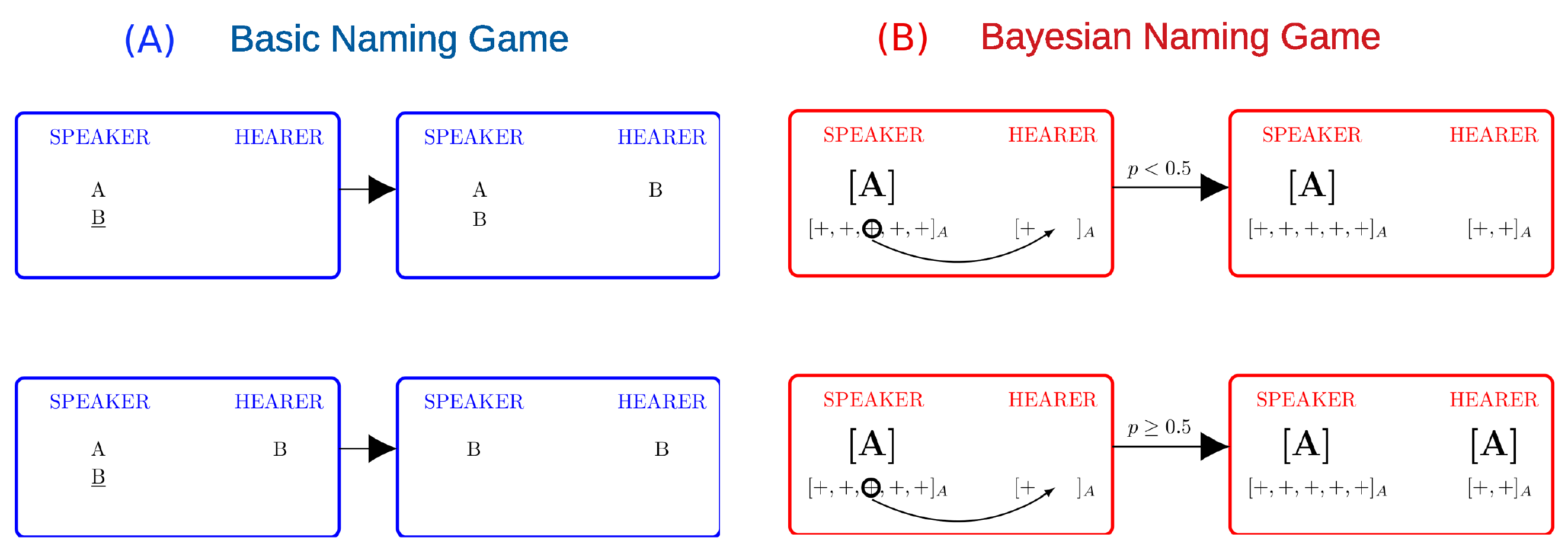}
\caption{
Comparison of the basic and Bayesian NG model.\\
Panel (A): basic 2-conventions NG model.
In a communication failure (upper figure), the name conveyed, $B$ in the example, is not present in the list of the hearer, who adds it to the list.
In a communication success (lower figure), the word $B$ is already present in the hearer's list and both agents erase $A$ from their lists.\\
Panel (B): Bayesian NG model; in order to convey an example ``+'' to the hearer in association with name $A$, the speaker must have already generalized concept $C$ in association with $A$, represented here by the label $[\mathbf{A}]$.
In a communication failure (upper figure), the hearer computes the Bayes probability $p$ and the result is a $p < 1/2$; then the only outcome is that the hearer records the example (\textit{reinforcement}).
In the Bayesian NG, there are two ways, in which the communication can be successful.
The first way (lower figure) is when $p \geq 1/2$: the hearer \textit{generalizes} $C$ in association with $A$ and attaches the label $[\textbf{A}]$ to the inventory.
The second way (not shown) is the the \textit{agreement} process, analogous to that of the basic NG, when both agents had already generalized concept $C$ in association with name $A$ and remove label $[\mathbf{B}]$ from the inventory $[+++\dots]_B$ if present. 
See text for further details.
}
\label{fig:cartoonNG1}
\end{center}
\end{figure}

The system evolves according to the following update rules \cite{Baronchelli-2016a}:
\begin{enumerate}
    \item 
    Two agents $i$ and $j$, the speaker and the hearer, respectively, are randomly selected.
    \item 
    The speaker $i$ randomly extracts a name (here either $A$ or $B$) from the list $\L_i$ and conveys it to the hearer $j$. 
    Depending on the state of agent $j$, the communication is usually described as:
    \begin{enumerate}
        \item 
        \emph{Success}: the conveyed name is present also in the hearer's list $\L_j$, i.e. also agent $j$ knows its  meaning; then the two agents erase the other name from their lists, if present.
        \item 
        \emph{Failure}: the conveyed name is \emph{not} present in the hearer's list $\L_j$; then agent $j$ records and adds it to the list $\L_j$.
    \end{enumerate}
\item 
Time is increased of one step, $t \to t + 1$,  and the simulation is reiterated from the first point above.
\end{enumerate}
An example of unsuccessful and one of successful communication are schematized in the left panel (A) of Fig.~\ref{fig:cartoonNG1}, see Ref. \cite{Baronchelli-2006c} for more examples.
Despite its simple structure, the basic NG model describes the emergence of consensus about which name to use, which is reached for any (disordered) initial configuration \cite{baronchelli2007}.

\subsection{Toward a Bayesian naming game model}

From a cognitive perspective, a ``communication failure'' of the NG model can be understood as a learning process, in which the hearer learns a new word.
It is a ``one-shot learning process'', because it takes place instantaneously (in a single time step) and independently of the the agent's history (i.e. of the previous knowledge of the agent).
However, modeling an actual learning process should take into account the agents' experience, based on the previous observations (the data already acquired) as well as the uncertain/incomplete character naturally accompanying any learning process.

Here, the one-shot learning is replaced by a process that can describe basic but realistic situations, such as the prototypical ``linguistic games'' \cite{Wittgenstein1953}.
For example, consider a ``lecture game'', in which a lecturer (speaker) utters the name $A$ of an object and shows a real example ``+'' of the object to a student (hearer), repeating this process a few times.
Then, the teacher can e.g. (a) show another example and ask the student to name the object; (b) utter the same name and ask the student to show an example of that object; or (c) do both things (uttering the name and showing the object) and ask the student whether the name$\leftrightarrow$object correspondence is correct.
The student will not be able to answer correctly if not after having received some examples, enabling the student to generalize the concept $C$ corresponding to the object in association to name $A$.
To model these and similar learning processes, we need a criterion enabling the hearer to assess the degree of equivalence between the new example and a the examples recorded previously.

The starting point for the replacement of the one-shot learning is Bayes' theorem.
According to Bayes' theorem, the posterior probability $p\left(h|X\right)$ that the generic hypothesis $h$ is the true hypothesis, after observing a new evidence $X$, reads \cite{Harney-2003a,Jeffreys1961},
\begin{equation}\label{eq:bayesTheorem}
p\left(h|X\right) = \frac{p\left(X|h\right) p\left(h\right)}{p\left(X\right)} \, .
\end{equation}
Here, the prior probability $p\left(h\right)$ gives the probability of occurrence of the hypothesis $h$ before observing the data and $p\left(X|h\right)$ gives the probability of observing $X$ if  $h$ is given.
Finally, $p\left(X\right)$ gives the normalization constraint; in the applications it can be evaluated as $p\left(X\right) = \sum_{h'} p\left(X|h'\right) p\left(h'\right)$, where $\{h'\} \in H$ represents the set of hypotheses, within the hypothesis space $H$.

\begin{figure}[h!]
\begin{center}
\includegraphics*[width=7cm]{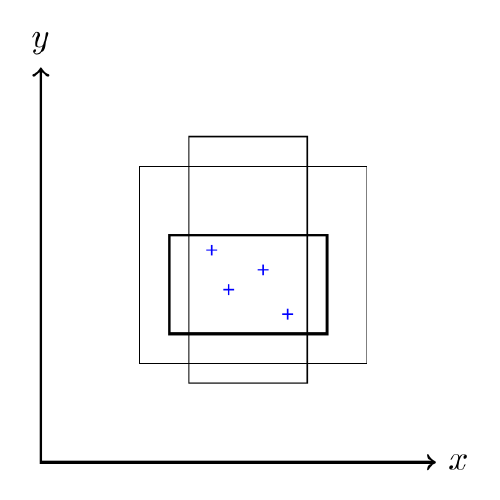}
\end{center}
\caption{
Three different hypotheses represented as axis-parallel rectangles in $\mathbb{R}^{2}$ and four positive examples ``+'' that are all consistent with the three hypotheses. 
The set of all the rectangles that can be drawn in the plane constitutes the hypothesis space $H$.}
\label{fig:cartoon2}
\end{figure}

The next step is to find a way to compute explicitly the posterior probability $p\left(h|X\right)$, through a representation of the concepts and their relative examples in a suitable hypothesis space $H$ of the possible extensions of a given concept $C$, constituted by the mutually exclusive and exhaustive hypotheses $h$.
Following the experimentally verified Bayesian statistical framework of Refs. \cite{Tenenbaum-1999,Tenenbaum-1999b}, we adopt the paradigmatic representation of a concept as a geometrical shape.
For example, the concept of ``healthy level'' of an individual in terms of the levels of cholesterol $x$ and insulin $y$, defined by the ranges $x_a \leq x \leq x_b$ and $y_a \leq y \leq y_b$, where $x_i$ and $y_i$ ($i = a,b$) are suitable values, represents a rectangle in the Euclidean $x$-$y$ plane $\mathbb{R}^{2}$.
Examples of healthy levels of specific individuals $1,2,\dots$ correspond to points $(x_1,y_1), (x_2,y_2), \dots \in \mathbb{R}^{2}$.
In the following, we assume that a hypothesis $h$ is represented by a rectangular region in $\mathbb{R}^{2}$.
Figure~\ref{fig:cartoon2} shows four positive examples, denoted by the symbol ``+'', associated to four different points of the plane, consistent with  (i.e. contained in) three different hypotheses, shown as rectangles.

The problem of learning a word is now recast into an equivalent problem, consisting in acquiring the ability to infer whether a new example $z$ recorded, corresponding to a new point ``+'' in $\mathbb{R}^{2}$,  corresponds to the concept $C$, after having seen a small set of positive examples ``+'' of $C$. 
More precisely, let $X = \left \{   \left(x_1,y_1 \right),  \dots,  \left(x_n,y_n \right)   \right \} $ be a sequence of $n$ examples of the true concept $C$, already observed by the hearer, and $z = (z_1,z_2)$ the new example.
The learner does not know the true concept $C$, i.e. the exact shape of the rectangle associated to $C$, but can compute the generalization function $p\left(z\in C |X\right)$ by integrating the predictions of all hypotheses $h$, weighted by their posterior probabilities $p\left(h|X\right)$: 
\begin{equation}\label{eq:generalization}
p \left(z \in C | X \right) = \int_{h \in H}p \left(z \in C | h \right) p\left(h|X\right) dh \, .
\end{equation}
Clearly, $p \left(z \in C | X \right)= 1 $ if $z \in  h$ and $0$ otherwise. 
By means of the Bayes' theorem \eqref{eq:bayesTheorem}, one can obtain the right Bayesian probability for the problem at hand.
A successful generalization is then defined quantitatively by introducing a threshold $p^*$, representing an acceptance probability:
\textit{an agent will generalize if the Bayesian probability $p \left(z \in C | X \right) \ge p^*$. }
The value $p^* = 1/2$ is assumed, as in Ref. \cite{Tenenbaum-1999b}.

We assume that an Erlang prior characterizes the agents' background knowledge. 
For a rectangle in $\mathbb{R}^{2}$ defined by the tuple $\left( l_1, l_2,  s_1,  s_2 \right)$, where  $l_1,l_2$ are the Cartesian coordinates of its lower-left corner and $s_i$ its sides along dimension $i=1,2$, the Erlang prior density is \cite{Tenenbaum-1999,Tenenbaum-1999b} 
\begin{equation}\label{eq:erlang}
p_E = s_1 s_2  \exp \left \{ -  \left(   \frac{s_1 }{\sigma_1} +     \frac{s_2 }{\sigma_2 }      \right)   \right \} \, ,
\end{equation}
where the parameters $\sigma_i$ represent the actual sizes of the concept, i.e. they are the sides of the concept rectangle $C$ along dimension $i$.
The choice of a specific informative prior, such as the Erlang prior,
is well motivated by the fact that in the real world individuals have always some prior knowledge or expectation.
In fact, a  Bayesian learning framework  with an Erlang prior of the form  \eqref{eq:erlang} well describes experimental observations of learning processes of human beings \cite{Tenenbaum-1999b}.
The final expression used below for computing the Bayesian probability $p$  that, given the set of previous examples $X$, the new example $z$ falls in the same category of concept $C$, reads \cite{Tenenbaum-1999b}
\begin{equation}\label{eq:one}
p \left(z \in C | X \right) 
  \approx 
  \frac{\exp \left \{ -  \left(   \frac{\tilde{d}_1 }{\sigma_1} +     \frac{\tilde{d}_2 }{\sigma_2 }      \right)   \right \} }
              {\left[ \left( 1 + \frac{\tilde{d}_1 }{r_1} \right) \left( 1 + \frac{\tilde{d}_2 }{r_2} \right) \right]^{n-2} }  \, .
\end{equation}
Here $r_i$ ($i = 1, 2$) is an estimate of the extension of the set of examples along direction $i$, given by the maximum mutual distance along dimension $i$ between the examples of $X$; 
$\tilde{d}_i$ measures an effective distance between the new example $z$ and the previously recorded examples, i.e., $ \tilde{d}_i = 0$ if $z_i$ falls inside the value range of the examples of $X$ along dimension $i$, otherwise $ \tilde{d}_i$ is the distance between $z$ and the nearest example in $X$ along the dimension  $i$.   
Equation~\eqref{eq:one} is actually a ``quick-and-dirty'' approximation that is reasonably good, except  for $ n \leq 3$ and  $ r_i \leq \sigma/10$, estimating the actual generalization function within a $10  \%$ error, see Refs. \cite{Tenenbaum-1999,Tenenbaum-1999b} for details.
Despite these approximations, Eq.~\eqref{eq:one}  will ensure that our computational model, described in the next section, retains the main features of the  Bayesian learning framework.
It is to be noticed that for the validity of the Bayesian framework, it is crucial that the examples are drawn randomly from the concept (strong sampling assumption), i.e. they are extracted from a probability density that is uniform in the rectangle corresponding to the true concept \cite{Tenenbaum-1999b}.
This definition of generalization is now applied below to word-learning.

\subsection{The Bayesian word-learning model}
\label{sec:computational0}

Based on the Bayesian learning framework discussed above, in this section we introduce a minimal Bayesian individual-based model of word-learning.
For the sake of clarity, in analogy with the basic NG model, we study the emergence of consensus in the simple situation, in which two names $A$ and $B$ can be used for referring to the same concept $C$ in pair-wise interactions among $N$ agents.

At variance with the NG model, here in each basic pair-wise interaction an agent $i$, acting as a speaker, conveys an example ``+'' of concept $C$, in association with either name $A$ or $B$, to another agent $j$, who acts as hearer ($i, j = 1, \dots, N$).
In order to be able to communicate concept $C$ uttering a name, e.g. name $A$, the speaker $i$ must have already generalized concept $C$ in association with name $A$. This is signalled by the presence of name $A$ in the list $\L_i$.
On the other hand, the hearer $j$ always records the example received in the respective inventory, in the example the inventory $[+++\dots]_A$.

The state of a generic agent $i$ at time $t$ is defined by 
\begin{itemize}

    \item the list $\L_i$, to which a name is added whenever agent $i$ generalizes concept $C$ in association with that name; agent $i$ can use any name in $\L_i$ to communicate $C$;

    \item two inventories $[+++\dots]_A$ and $[+++\dots]_B$, containing the examples ``+'' of concept $C$ received from the other agents in association with name $A$ and $B$, respectively.

\end{itemize}

It is assumed that initially each agent knows one word: a fraction $n_A(0)$ of the agents know concept $C$ in association with name $A$ and the remaining fraction $n_B(0) = 1 - n_A(0)$ in association with name $B$ --- no agent knows both words, $n_{AB}(0) = 0$.
We will examine three different initial conditions:
\begin{align}
   \textrm{symmetric initial conditions}~
   & \textrm{(\textbf{SIC}):} 
   & n_A(0) & = n_B(0) = 0.5 \nonumber\\
   \textrm{asymmetric initial conditions}~
   & \textrm{(\textbf{AIC}):} 
   & n_A(0) & = 0.3,~~~ n_B(0) = 0.7 \nonumber\\
  \textrm{reversed case of AIC}~
  & \textrm{(\textbf{AICr}):}
  & n_A(0) & = 0.7,~~~ n_B(0) = 0.3 &  \nonumber
\end{align}
Initially, each agent $i$, within the fraction $n_A(0)$ of agents that know name $A$, is assigned $n_{ex, A} = 4$ examples ``+'' of concept $C$ in association with name $A$, but no examples in association with the other name $B$, so that agent $i$ has an $A$-inventory $[+ + + +]_A$ and an empty $B$-inventory $[ \cdot ]_B$.
The complementary situation holds for the other agents that know only name $B$, who initially receive $n_{ex, B} = 4$ examples of concept $C$ in association with name $B$ but none in association with $A$.
This choice, somehow arbitrary, is dictated by the condition that Eq.~\eqref{eq:one} becomes a good approximation for $n>3$  \cite{Tenenbaum-1999}.

Examples are points uniformly generated inside the fixed rectangle corresponding to the true concept $C$, here assumed to be a rectangle with lower left corner coordinates $(0,0)$ and sizes $\sigma_1=3$ and $\sigma_2=1$ along the $x$ and $y$ axis, respectively.
Results are independent of the assumed numerical values; in particular, no appreciable variation in the convergence times  $t_{conv}$ is observed as the rectangle area is varied, which is consistent with  the strong sampling assumption, on which the Bayesian learning framework rests; 
see Ref. \cite{Tenenbaum-1999} and Sec. \ref{sec:results}.

Furthermore, we introduce an element of asymmetry between the names $A$ and $B$, related to the word-learning process: different \textit{minimum numbers of examples} $n^{\ast}_{ex, A} = 5$ and $n^{\ast}_{ex, B} = 6$ will be used, which are needed by agents to generalize concept $C$ in association with $A$ and $B$, respectively.
This is equivalent to assume that concept $C$ is slightly easier to learn in association with name $A$ than $B$.
Such an asymmetry plays a relevant role in the model dynamics in differentiating the Bayesian generalization functions $p_A$ and $p_B$ from each other, see Sec. \ref{analysis}.

The dynamics of the model can be summarized by the following update rules:
\begin{enumerate}

\item A pair of agents $i$ and $j$, acting as speaker and hearer,  respectively, are randomly chosen among the agents.

\item The speaker selects randomly: 
(a) a name from the list $\L_i$ (or selects the name present if $\L_i$ contains a single name), for example $A$ (analogous steps follow if the word $B$ is selected);
(b) an example $z$ among those contained in the corresponding inventory $[+++\dots]_A$ --- ;\\ 
then the speaker $i$ conveys the example extracted $z$ in association with (e.g. uttering) the name selected $A$ to the hearer $j$. 

\item The hearer adds the new example $z$ (in association with $A$) to the inventory $[+++\dots]_A$. This \emph{reinforcement} process of the hearer's knowledge always takes place.

\item Instead, the next step depends on the state of the hearer:

	\begin{enumerate}

		\item \emph{Generalization}. If the selected name, $A$ in the example, is \emph{not} present in the hearer's list $\L_j$, then the hearer $j$ computes the relative Bayesian probability $p_A = p(z \in C | X_A)$ that the new example $z$ falls in the same category of concept $C$, using the examples previously recorded in association with $A$, i.e. from the set of examples $X_A \in [+++\dots]_A$. 
		If $p_A \ge 1/2$, the hearer has managed to generalize concept $C$ and connects the inventory $[+++\dots]_A$ to name $A$; this is done by adding name $A$ to the list $\L_j$.
		Starting from this moment, agent $j$ can communicate concept $C$ to other agents by conveying an example taken from the inventory $[+++\dots]_A$ while uttering the name $A$.	
		If $p_A < 1/2$, the hearer has not managed to generalize the concept and nothing more happens (the reinforcement of the previous point is the only event taking place).

		\item  \emph{Agreement}. The name uttered by the speaker, $A$ in the example, is present in the hearer's list $\L_j$, meaning that that agent $j$ has already generalized concept $C$ in association with name $A$ and has connected the corresponding inventory $[+++\dots]_A$ to $A$.
		In this case, the hearer and the speaker proceed to make an agreement --- analogous to that of the NG model, leaving $A$ in their lists $\L_i$ and $\L_j$ and removing $B$ is present.
		No examples contained in any inventory are removed.
		
		\end{enumerate}
\item Time is updated, $t \to t + 1$, and the simulation is reiterated from the first point above.

\end{enumerate}

Two examples of Bayesian word-learning process, a successful and an unsuccessful one, are illustrated in the cartoon in the right panel (B) of Fig. \ref{fig:cartoonNG1}.
Table \ref{table:interactions} lists the possible encounter situations, together with the corresponding relevant probabilities.

Notice that an agent $i$ can enter a pair-wise interaction with a non-empty inventory of examples, e.g. $[+++\dots]_A$, associated to name $A$, without being able to use name $A$ to convey examples to other agents, i.e., without the name $A$ in the list $\L_i$ due to not having generalized concept $C$ in association with $A$.
Those examples can have different origins: 
(1) in the initial conditions, when $n_{ex,A}$ randomly extracted examples associated to $A$ and $n_{ex,B}$ to $B$ are assigned to each agent;
(2) in previous interactions, in which the examples were conveyed by other agents;
(3) in an agreement about convention $B$, which removed label $A$ from the list $\L_i$ while leaving all the corresponding examples in the inventory associated to name $A$.
In the latter case, the inventory $[+++\dots]_A$ may be ``ready'' for a generalization process, since it contains a sufficient number of examples, i.e., agent $i$ will probably be able to generalize as soon as another example is conveyed by an agent.
This situation is not as peculiar as it may look at first sight.
In fact, there is a linguistic analogue in the case where a speaker that loses the habit to use a certain word (or a language) $A$ can regain it promptly, if exposed to $A$ again.

Notice also that without the agreement dynamics scheme introduced in the model, borrowed from the basic NG model, the population fraction $n_{AB}$ of individuals who know both $A$ and $B$ ($n_{A} + n_{B} + n_{AB}= 1$) would be growing, until eventually $n_{AB} = 1$.

\begin{table}
\small
\centering
\caption{\small\label{table:interactions}
Pair-wise interactions in the Bayesian NG model.
The speaker (S) conveys a name $\arrA$ or $\arrB$ to the hearer (H) together with an example taken from the speaker's inventory, $[+++\dots]_A$ or $[+++\dots]_B$, respectively --- this happens with a branching probability $q = 0.5$ if the speaker has the list $(A,B)$ and knows the meaning of both names.
The outcome can be: 
(1) a \textit{reinforcement} (only); 
(2) \textit{generalization} of concept $C$, if the Bayes probability is $p > 1/2$;
(3) an \textit{agreement} between hearer and speaker, if both agents know the meaning of the conveyed name. 
Even if not indicated, reinforcement takes place also in cases (2) and (3).
} 
\def\arraystretch{1.5}
\begin{tabular}{| l   l   l   l   l   l   l   l | }
\hline 
{S-List}        & { Name}       & { H-List}     & { Branching}      & { Process}    & Condition     & {S- List} & { H-List}  \\ 
{ (before)}     & { conveyed}   & { (before)}   & { probability}    &   ~ 			&  ~            & {(after)} & { (after)} \\
\hline
$(A)$  	        & $\arrA$       & $(A)$         & $(q=1)$           & Reinforcement & always        & $(A)$ 	    & $(A)$ \\ \hline
$(A)$  	        & $\arrA$       & $(B)$         & $(q=1)$           & Reinforcement & $p_A < 1/2$   & $(A)$ 	    & $(B)$ \\ 
~               & ~             & ~             & $(q=1)$           & Learning      & $p_A \ge 1/2$ & $(A)$ 	    & $(A,B)$ \\ \hline
$(A)$  	        & $\arrA$       & $(A,B)$       & $(q=1)$           & Agreement     & always        & $(A)$ 	    & $(A)$ \\ \hline
$(B)$  	        & $\arrB$       & $(A)$         & $(q=1)$           & Reinforcement & $p_B < 1/2$   & $(B)$ 	    & $(A)$ \\ 
~   	        & ~		        & ~             & $(q=1)$           & Learning      & $p_B \ge 1/2$ & $(B)$ 	    & $(A,B)$ \\ \hline
$(B)$	        & $\arrB$       & $(B)$     	& $(q=1)$           & Reinforcement & always		& $(B)$ 	    & $(B)$ \\ \hline
$(B)$  	        & $\arrB$       & $(A,B)$       & $(q=1)$           & Agreement     & always 		& $(B)$ 	    & $(B)$ \\ \hline
$(A,B)$         & $\arrA$       & $(A)$	        & $q=1/2$           & Agreement		& always		& $(A)$ 	    & $(A)$ \\ \hline
$(A,B)$         & $\arrB$       & $(A)$         & $q=1/2$           & Reinforcement & $p_B < 1/2$   & $(A,B)$ 	    & $(A)$ \\ 
~  		        & ~  	        & ~		        & ~                 & Learning      & $p_B \ge 1/2$ & $(A,B)$ 	    & $(A,B)$ \\ \hline
$(A,B)$	        & $\arrA$       & $(B)$         & $q=1/2$           & Reinforcement & $p_A < 1/2$   & $(A,B)$ 	    & $(B)$ \\ 
~  		        & ~  	        & ~		        & ~                 & Learning      & $p_A \ge 1/2$ & $(A,B)$ 	    & $(A,B)$ \\ \hline
$(A,B)$	        & $\arrB$       & $(B)$         & $q=1/2$           & Agreement     & always		& $(B)$ 	    & $(B)$ \\ \hline
$(A,B)$	        & $\arrA$       & $(A,B)$       & $q=1/2$           & Agreement	 	& always		& $(A)$ 	    & $(A)$ \\ \hline
$(A,B)$	        & $\arrB$       & $(A,B)$       & $q=1/2$           & Agreement	 	& always		& $(B)$ 	    & $(B)$ \\ \hline
\end{tabular}
\end{table}


\section{Results}
\label{sec:results}

In this section we study  numerically the Bayesian NG model introduced above and discuss its main features.
We limit ourselves to  study the model dynamics on a fully-connected network.

\begin{figure}[h!]
\begin{center}
\includegraphics[width=8cm]{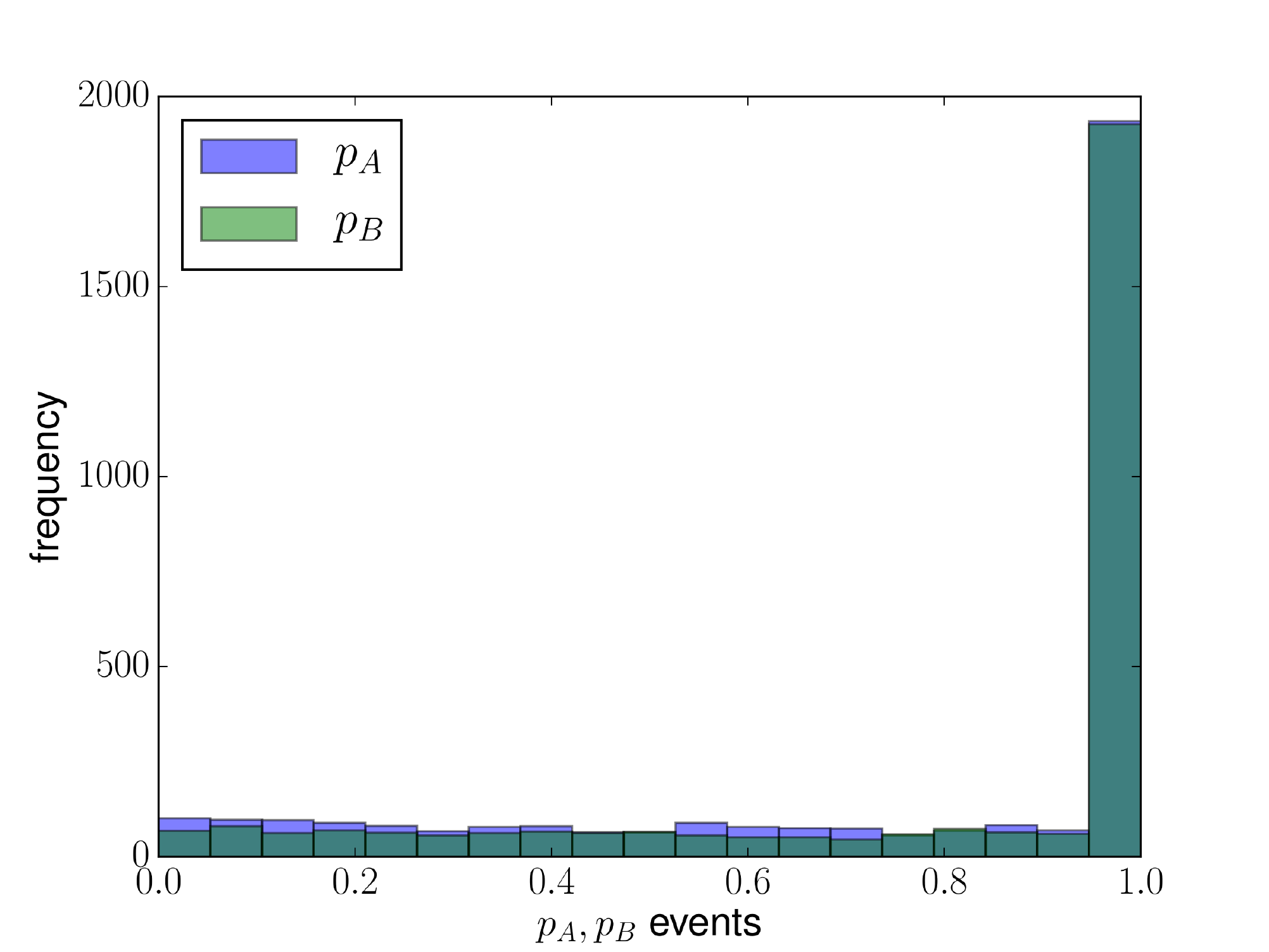}
\end{center}
\caption{Histograms of the Bayesian probabilities $p_A, p_B$ computed by agents during their learning attempts during a single run (for $N=2000$ agents, starting with SIC; $n^{\ast}_{ex,A}=5$,  $n^{\ast}_{ex,B}= 6$).}
\label{fig:histogram} 
\end{figure}

In the new learning scheme, which replaces the one-shot learning of the two-conventions NG model, an individual generalizes concept $C$ on a suitable time scale $\Delta t > 1$, rather than during a single interaction.
However, a few examples are sufficient for an agent to generalize concept $C$, as in a realistic concept-learning process.
This is visible from the Bayesian probabilities $p_A$ and $p_B$ computed by agents in the role of hearer, according to Eq.~\eqref{eq:one}, once at least  $n^{\ast}_{ex, A}=5$ and $n^{\ast}_{ex, B}=6$ examples  ``+'', respectively,  have been  stored in the inventory associated to the name $A$ and $B$: Figure \ref{fig:histogram} shows the histograms of the $p_A$'s and $p_B$'s computed from the initial time until consensus for a single run with $N=2000$ agents and starting with SIC.
The low frequencies at small values of $p_{A}$ and $p_{B}$ and the highest frequencies at values close to unity are due to the fact that the Bayesian probabilities reach values $p_{A} \approx p_{B} \approx 1$ very fast, after a few learning attempts, consistently with the size principle, on which the Bayesian learning paradigm, and in turn Eq.~\eqref{eq:one}, are based \cite{Tenenbaum-1999}. 

\begin{figure}[h!]
\begin{center}
\includegraphics*[width=17cm]{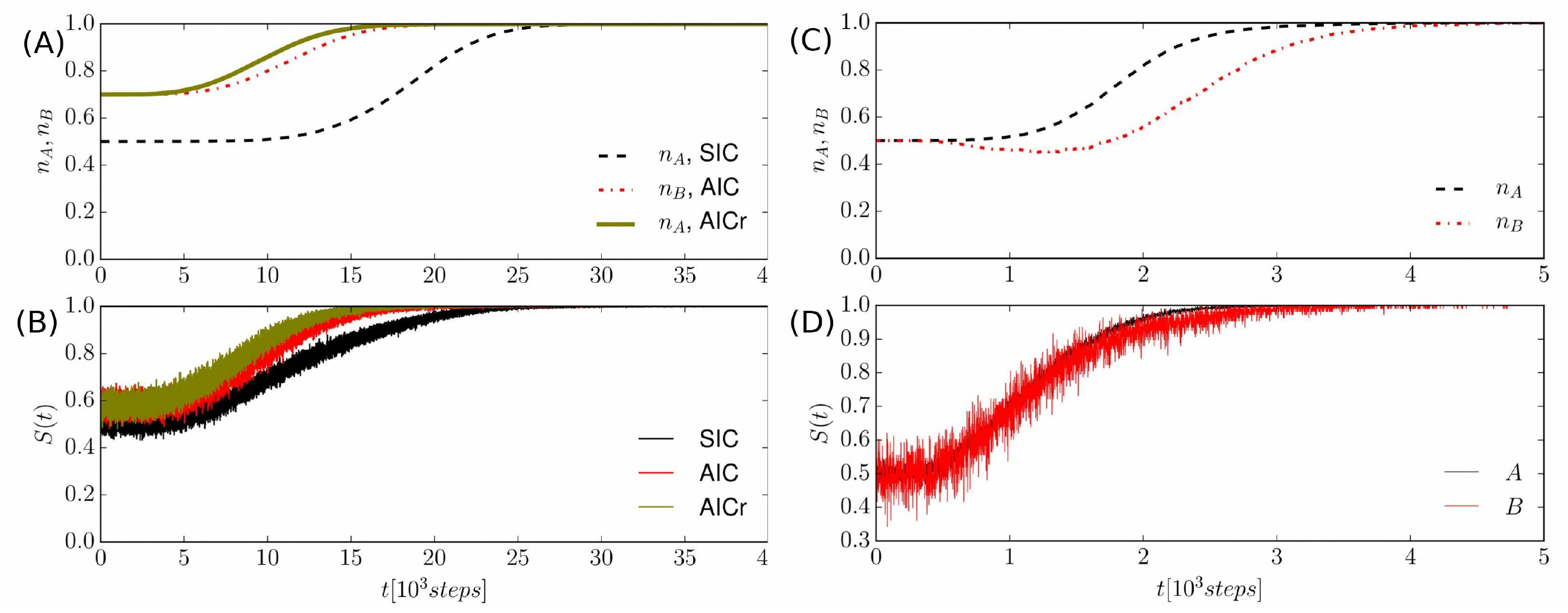}
\end{center}
\caption{
Average population fraction associated to the name shared in the final consensus state (upper panels (A) and (C)) and success rate $S(t)$ (lower panels (B) and (D)) \textit{versus} time.
Left panels (A) and (B): system with $N = 1000$ agents starting from different initial conditions, SIC, AIC, and AICr; averages done over $600$ runs.
Right panels (C) and (D):  system with $N = 100$ agents starting from SIC; averages done over $1000$ runs --- notice that due to the smaller size $N=100$, the system can converges to consensus both with name $A$ (in a fraction of cases $p_{e,A } \approx 0.9)$ and with name $B$ ($p_{e,B } \approx 0.1$).
} 
\label{fig:model2} 
\end{figure}

In order to visualize how the system approaches consensus, it is useful to consider some global observables, such as the fractions $n_A(t)$, $n_B(t)$, and $n_{AB}(t)$ of agents that have generalized concept $C$ in association with name $A$ only, name $B$ only, or both names $A$ and $B$, respectively, or the success rate $S(t)$.
The dynamics of a population of $N=1000$ agents (panels (A) and (B)) using different initial conditions, SIC, AIC, and AICr, and that of a population of $N=100$ agents starting with SIC (panels (C) and (D)) are shown in Fig. \ref{fig:model2}. 

Panel (A) of Fig. \ref{fig:model2} shows only the population fractions corresponding to the name found at consensus, for the sake of clarity (the remaining population fractions eventually go to zero).
For asymmetrical initial condition (AIC or AICr), it is the initial majority that determines the convention found at consensus (that is $B$ for AIC and $A$ for AICr). 
If the system starts from SIC, the convention $A$, for which agents can generalize earlier ($n^{\ast}_{ex, A}=5  < n^{\ast}_{ex, B} = 6$), is always found at consensus --- in this case it is the asymmetry in the thresholds $n^{\ast}_{ex, A}$ and $n^{\ast}_{ex,B}$, characterizing the Bayesian learning process, to determine consensus.

Panel (B) of Fig. \ref{fig:model2} shows the success rate $S(t=t_k)$, representing the average over different runs of the instantaneous success rate $S_k$ of the $k$th interaction at time $t_k$, defined as follows:
$S_k = 1$ in case of agreement between the two agents or when a successful learning of the hearer takes places, following a Bayes probability $p > 1/2$; or $S_k = 0$ in case of unsuccessful generalization, when $p < 1/2$ and only reinforcement takes place.
The success rate $S(t)$ varies between $S(0) \approx (n_A(0))^2 + (n_B(0))^2$, due to the respective fractions of agents that initially know the two conventions $A$ and $B$, to $S \approx 1$ at consensus, following a typical S-shaped curve of learning processes \cite{Baronchelli2006}.
In the case of SIC, the initial value is $S(0) \approx 0.5^2+0.5^2=0.5$, while for AIC or AICr the initial value is $S(0) \approx (0.3)^2 + (0.7)^2 \approx 0.58$.

\begin{figure}
\begin{center}
\includegraphics*[width=16cm]{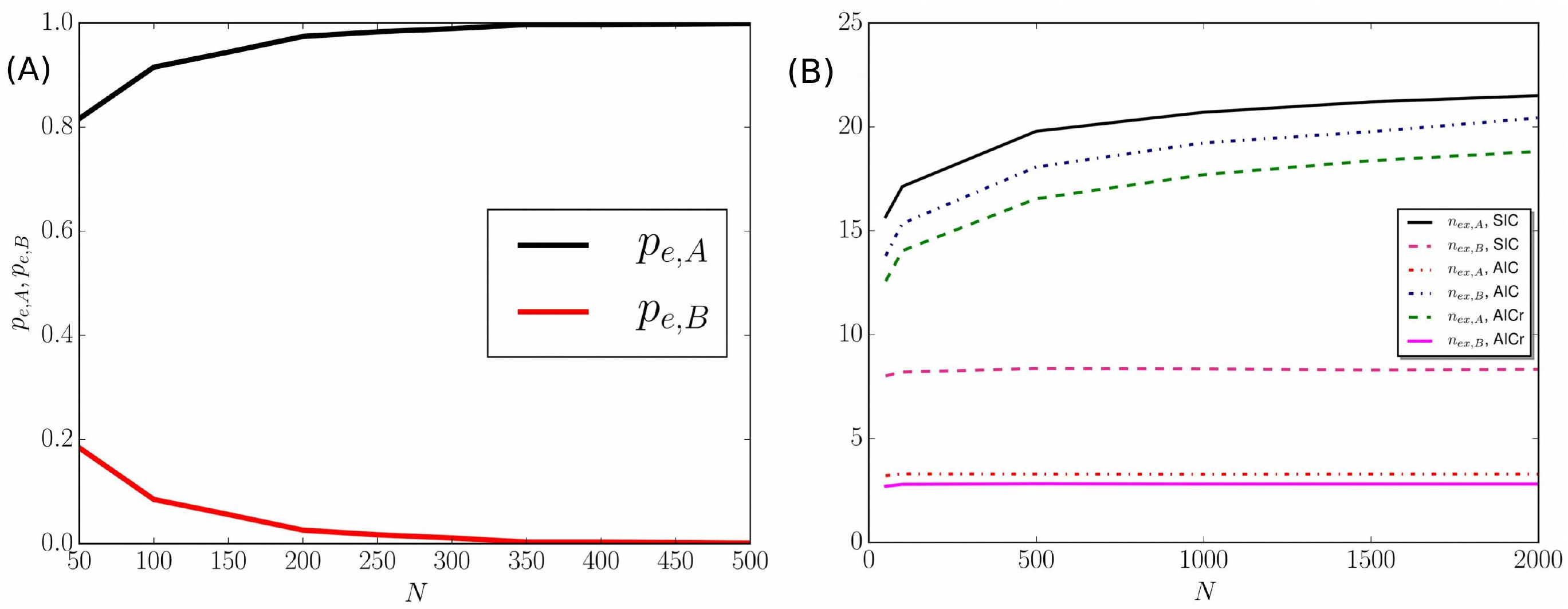}
\end{center}
\caption{Panel (A): probabilities $p_{e,A }$ and $p_{e,B }$ that the system reaches the consensus at $A$ and $B$ respectively, \textit{versus} the system sizes $N$, obtained by averaging over $1000$ runs of a system starting with SIC. 
Panel (B): average number of examples  $\bar{n}_{ex, A}$ and $\bar{n}_{ex, B}$ recorded by an agent at consensus, for a system of $N= 50, 100, 500,1000, 1500, 2000$ agents, starting with SIC, AIC, AICr. Averages are done over $600$ runs.
}
\label{fig:rate}
\end{figure}

We now investigate how the modified Bayesian dynamics affects the convergence times to consensus.
The study of the size-dependence of the convergence to consensus shows that there is a critical  value  $ N^{\ast}  \approx 500$  in the case of SIC, such that for $N \leq  N^{\ast}$ there is a  non-negligible probability that the final absorbing state is $B$. 
Panels (C) and (D) of Fig. \ref{fig:model2}, representing the results for a system starting with SIC and a smaller size $N=100$, show the existence of two possible final absorbing states and that there are different times scales associated to the convergence to consensus:
name $A$ is found at consensus in about $90\%$ of cases and name $B$ in the remaining cases.
The branching probability into $A$ or $B$ consensus is further investigated in panel (A) of Fig. \ref{fig:rate}, where we plot the branching probabilities $p_{e,A }, p_{e,B }$ \textit{versus} the system sizes $N$.
The nonlinear behavior (symmetrical sigmoid) signals the presence of finite-size effects, particularly clear for relatively small $N$-values. 
In fact, when the fluctuations in the system are larger, the system size can play an important role in the dynamics of social systems, as an actual thermodynamic limit is only allowed for simulations of macroscopic physical systems  \cite{Toral2007a}.

The convergence time $t_{conv}$  follows a simple scaling rule with the system size $N$, related to the average number of examples $\bar{n}_{ex, A}, \bar{n}_{ex, B}$ relative to $A,B$ respectively, stored  in the agents' inventories at consensus. 
These values depend on the number of learning and reinforcement processes, and hence are related to the system size $N$. 
The average number of interactions undergone by the agents until the system reaches the consensus is given by the sum $\bar{n}_{int}  = \bar{n}_{ex, A} + \bar{n}_{ex, B} $~\footnote{The $n_{ex, A} = n_{ex, B} = 4$ examples given initially to each agent are not accounted for by $\bar{n}_{ex, A}$ and $\bar{n}_{ex, B}$.}.
One expects that
\begin{equation}\label{eq:tconv}
t_{conv}  \approx  \bar{n}_{int} N  \, ,
\end{equation}
which suggests a linear scaling law  ($t_{conv} \sim N$) for convergence time with the system size $N$ for all the possible initial conditions. 
A linear behavior is indeed confirmed by the numerical simulations with population sizes  $N= 50, 100, 500,1000, 1500, 2000$ starting from SIC, AIC, AICr. The relative numerical results are reported in Table \ref{table:table2}. 
Moreover, in Eq. \eqref{eq:tconv}  the size-dependence of $\bar{n}_{int}$ is ignored as it shows a weak dependence upon $N$, see panel (B) in Fig. \ref{fig:rate}.

\begin{table}[h]
\renewcommand{\arraystretch}{2} 
\caption{Scaling laws $t_{conv} \sim N^{\alpha}$ with the system size $N$. Here the parameters are  $n^{\ast}_{ex,A}=5$,  $n^{\ast}_{ex,B}= 6$ with initial conditions SIC, AIC and AICr. The average number of examples, $\bar{n}_{ex, A}, \bar{n}_{ex, B}$, stored at  $t_{conv}$,  are obtained averaging over $600$ runs of a system with $N= 1000$ agents.
\label{table:table2}}
\centering
\begin{tabular}{c c c c  c}
\hline \hline
 & $\alpha$  & $\bar{n}_{ex, A}$  &  $\bar{n}_{ex, B}$ & outcome \\ [1ex]
\hline
SIC       & $1.06$          & $ 20$       & $8$   & $A, B$   \\
AIC       & $ 1.08$         & $3$         & $19$  & $B$   \\ 
AICr      & $1.09$          & $18$        & $3$   & $A$   \\ [1ex]
\hline \hline
\end{tabular}
\end{table}

From the above mentioned scaling law, it is clear that the average number of examples stored by the agents at consensus plays an important role in the semiotic dynamics. 
In particular, it is found that if the final absorbing state is $A$ (or B), then $\bar{n}_{ex, A} > \bar{n}_{ex, B}$ ( $\bar{n}_{ex, B} > \bar{n}_{ex, A}$). Moreover, the  average number of examples, relative to the absorbing state, always increases monotonically with the system size while a size-independent behavior is observed in the opposite case, see the right panel (B) of Fig. \ref{fig:rate}.

\begin{figure}[h!]
\begin{center}
\includegraphics*[width=10cm]{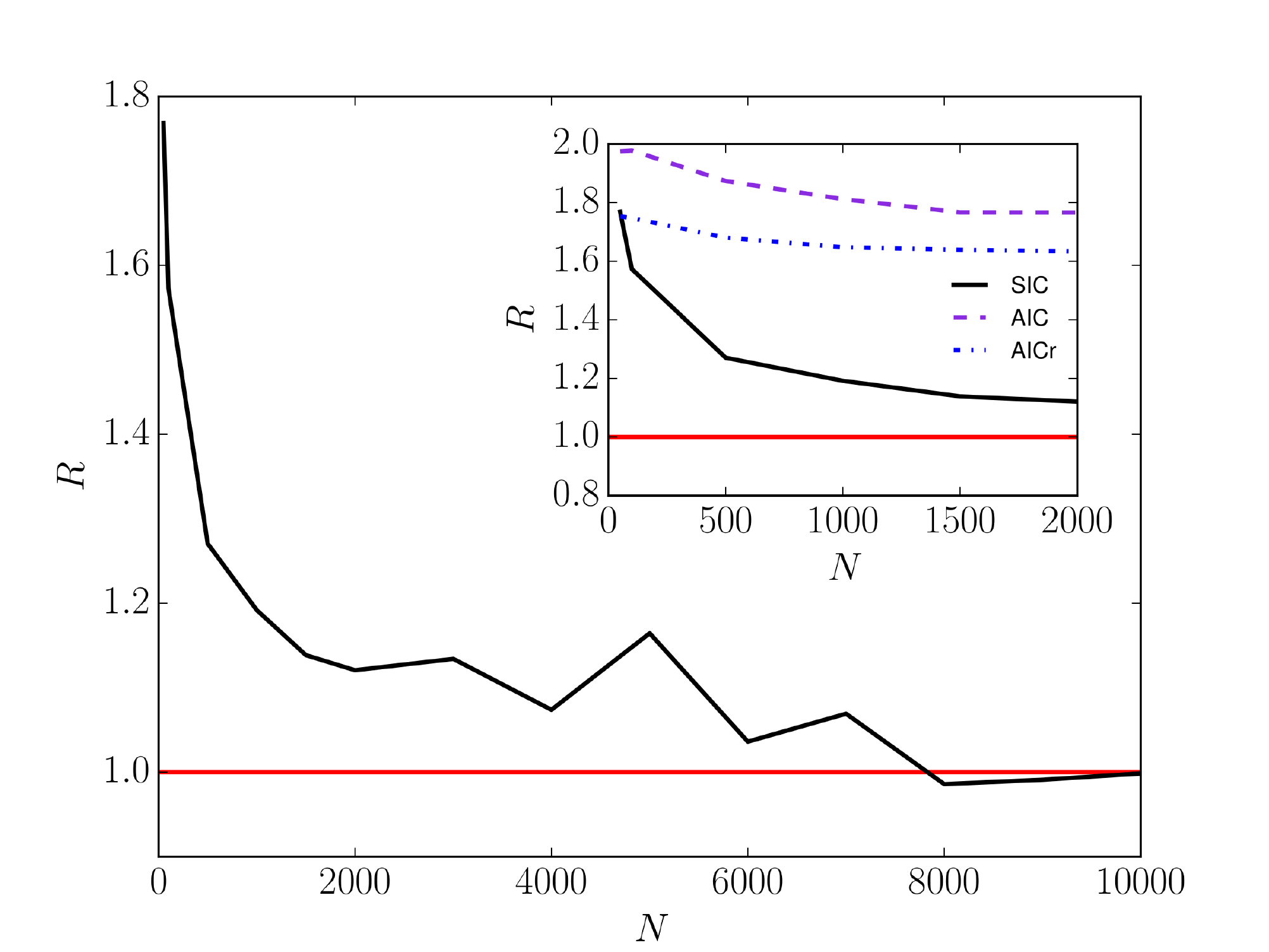}
\end{center}
\caption{The ratio of the convergence times of the Bayesian word-learning model and the 2-conventions NG model, $R = t_{conv}/\bar{t}_{conv}$, \textit{versus} the system size $N$, for a system starting with SIC. 
The inset illustrates the dependence of $R$ on different initial conditions. 
The curves are obtained averaging over  $900$ runs. }
\label{fig:convergence}
\end{figure}

Finally, we compare the convergence time of the Bayesian word-learning model, $t_{conv}$, with that of two-conventions NG model, $\bar{t}_{conv}$ \cite{Castello2009}, by studying the corresponding ratio $R = t_{conv}/\bar{t}_{conv}$ for common initial conditions and population sizes. 
When starting with SIC, the values of the convergence times obtained from the two models become of the same order by increasing $N$: $R$ decreases with $N$, reaching unity for $N=10000$, see Fig. \ref{fig:convergence}.
In other words, the time scales of the two models become equivalent for relatively large system sizes, i.e., the learning processes of the two models perform equivalently and the Bayesian approach roughly gives rise to the one-shot learning that characterizes the two-conventions NG model. 
In the next section we discuss how the Bayesian model becomes asymptotically equivalent to the minimal NG model.
The inset of Fig. \ref{fig:convergence} represents $R$ \textit{versus} $N$, for $N < 2000$, given different starting configurations, with SIC, AIC and AICr, and different population sizes.
In the following, we focus on the case of SIC.

\begin{figure}
\includegraphics*[scale=1.5]{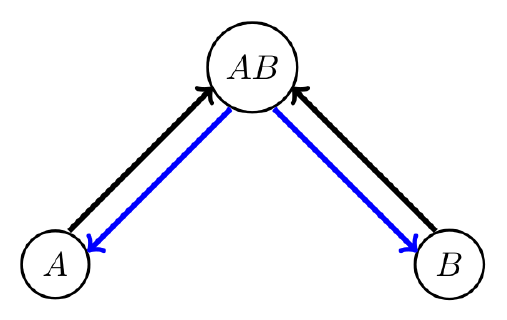}
\caption{Model scheme with two non-excluding options.
Arrows indicate allowed transitions between the ``bilingual'' state ($A$,$B$) and the ``monolingual'' states $A$ and $B$.
Direct $A \leftrightarrow B$ transitions are not allowed.
} 
\label{fig:diagram}
\end{figure}

\section{Stability analysis}
\label{analysis}

In this section we investigate the stability properties of the mean-field dynamics of the Bayesian NG model, in which statistical fluctuations and correlations are neglected.
In the Bayesian NG model, as in the basic NG, agents can use two non-excluding options $A$ and $B$ to refer to the same concept $C$.
The main difference between the Bayesian model and the basic NG model is in the learning process: a one-shot learning process in the basic NG and a Bayesian process in the Bayesian NG model.
In the latter case the presence of a name in the word list indicates that the agent has generalized the corresponding concept from a set of positive recorded examples. 

The NG model belongs to the wide class of models with two non-excluding options $A$ and $B$,  such as many models of bilingualism \cite{Patriarca2012a}, in which transitions between state $(A)$ and state $(B)$ are allowed only through an intermediate (``bilingual'') state $(A,B)$, as schematized in Fig. \ref{fig:diagram}.
The mean-field equations for the fractions $n_A(t)$ and $n_B(t)$ can be obtained considering the gain and loss contributions of the transitions depicted in Fig. \ref{fig:diagram},
\begin{align} \label{eq:lossgain1}
\dot{n}_A = p_{AB \rightarrow A^{\,}} \, n_{AB} - p_{A \rightarrow AB} \, n_A   \, , \nonumber \\
\dot{n}_B = p_{AB \rightarrow B} \, n_{AB} - p_{B \rightarrow AB} \, n_B \, .
\end{align}
Here $\dot{n_a}(t)=dn_a(t)/dt$ and the quantities $p_{a \rightarrow b}$ represent the respective transition rates per individual, corresponding to the arrows in Fig. \ref{fig:diagram} ($a,b = A, B, AB$). 
The equation for $n_{AB}(t)$ was omitted, since it is determined by the condition that the total number of agents is constant, $n_A(t) + n_B(t) + n_{AB}(t) = 1$.

The details of the possible pair-wise interactions in the Bayesian naming game are listed in Table \ref{table:interactions}.
From the various contributions, one obtains the master equation
\begin{align}\label{eq:autonomous}
  \dot{n}_A = - p_B n_A n_B +  n_{AB}^{2} + \frac{3 - p_B}{2} n_A n_{AB}  \, , \nonumber\\
  \dot{n}_B = - p_A n_A n_B +  n_{AB}^{2} + \frac{3 - p_A}{2} n_B n_{AB}  \, ,
\end{align}
which can be rewritten in the form \eqref{eq:lossgain1} with transition rates per individual given by
\begin{align}
\label{eq:learningP}
& p_{A \rightarrow AB} = p_B n_B + \frac{1}{2} p_B n_{AB} \, ,
\qquad 
& p_{B \rightarrow AB} = p_A n_A + \frac{1}{2} p_A n_{AB} \, ,
& ~\\
\label{eq:agreementP}
& p_{AB \rightarrow A} = \frac{3}{2}n_A +  n_{AB} \, ,
\qquad 
& p_{AB \rightarrow B} = \frac{3}{2} n_B +  n_{AB} \, .
& ~
\end{align}
Equations \eqref{eq:learningP} provide the transition rates of \textit{learning} processes, while Eqs. \eqref{eq:agreementP} give the transition rates of \textit{agreement} processes.
Setting $x \equiv n_A$, $y \equiv n_B$, and $z = n_{AB} \equiv 1 - x - y$, the autonomous system \eqref{eq:autonomous} becomes
\begin{align}\label{eq:ff}
\dot{x} = 
f_x\left(x,y\right) \equiv 
- p_B x y + (1 - x - y)^2 + \frac{1}{2}(3 - p_B)x(1 - x - y) \, ,\\
\dot{y} = 
f_y\left(x,y\right) \equiv 
- p_A x y + (1 - x - y)^2 + \frac{1}{2}(3 - p_A)y(1 - x - y) \, ,
\end{align}
where $\mathbf{v} = \left( f_x(x,y), f_y(x,y) \right)$ 
is the velocity field  in the phase plane.
For the following analysis, it is convenient to write the Bayesian probabilities $p_A$ and $p_B$ appearing in these equations as time-dependent parameters of the model, but they are actually highly non-linear functions of the variables.
In fact, they can be thought as averages of the microscopic Bayesian probability in Eq. \eqref{eq:one} over the possible dynamical realizations.
For this reason, they have also a complex non-local time-dependence on the previous history of the interactions between agents.
For the moment, we assume $p_A(t) = p_B(t) = p(t)$, returning later to the general case. 

From the conditions defining the critical points, 
$f_x\left(x,y\right) = f_y\left(x,y\right) = 0$, 
one obtains $\left(x - y\right) z = 0$.
Setting $z = 0$, one obtains two solutions that correspond to consensus in $A$ or $B$, given by 
$(x_1,y_1,z_1) = (1,0,0)$ and $(x_2,y_2,z_2) = (0,1,0)$.
Instead, setting $\left(x - y\right)= 0 $ leads to the equation
\begin{equation} \label{eq:secondOrd}
2 x^{2}  - (p + 5)x + 2 = 0 \, ,
\end{equation}
that has the solutions 
\begin{equation}
\label{eq:secondOrdSol}
x_\pm =  \frac{p+5 \pm \sqrt{(p+5)^{2} - 16}}{4} \, .
\end{equation}
For $p \in \left( 0, 1\right]$, the corresponding solutions $(x_\pm,x_\pm,1 - 2 x_\pm)$ are not suitable solutions, because $z_\pm = 1 - 2 x_\pm < 0$.

This analysis is valid for $p > 0$.
In fact, $p = p(t)$ is a function of time and for a finite interval of time after the initial time one has that $p=0$, which defines a different dynamical system.
In the initial conditions used,  $z(0) = 0$, which implies $z(t) = 0$, $x(t) = x(0)$, and $y(t) = y(0)$ at any later time $t$ as long as $p(t)=0$, since $\dot{x}(t) = \dot{y}(t) = \dot{z}(t) = 0$ (see Eq. \eqref{eq:autonomous};
in fact, the whole line $x + y = 1$ (for $0 < x,y < 1$) represents a continuous set of equilibrium points.
The reason why in this model $p(0)=0$ at $t=0$ and also during a subsequent finite interval of time is twofold.
First, agents do not have any examples associated to the name not known and they have to receive at least $n_{ex,A}^\ast$ or $n_{ex,B}^\ast$ examples, before being able to compute the corresponding Bayesian probability $p_A(t)$ or $p_B(t)$ --- thus it is to be expected that $p(t) = 0$ meanwhile.
Furthermore, even when agents can compute the Bayesian probabilities, the effective probability to generalize is actually zero, due to the threshold $p^\ast = 0.5$ for a generalization to take place.
The existence of the (temporary) equilibrium points on the line $x + y = 1$ ends as soon as the parameter $p(t) > p^\ast$ and, according to Eqs.~\eqref{eq:autonomous}, the two $A$- and $B$-consensus states become the only stable equilibrium points.
The representative point in the $x$-$y$-plane is deemed to leave the initial conditions on the $z = 1 - x - y = 0$ line, due to the stochastic  nature of the dynamics, which  is not invariant under time reversal \cite{hinrichsen2006}.

To determine the nature of the critical points $(x_1,y_1) = (1,0)$ and $(x_2,y_2) = (0,1)$, one needs to evaluate at the equilibrium points the $2 \times 2$ Jacobian matrix $A(x,y) = \{\partial_i f_j\}$, where $i,j = x,y$.
It is easy to show that both the critical points $(0,1)$ and $(1,0)$ are asymptotically stable \cite{strogatz2000}.

As long as the general case $p_A \ne p_B$, it can be shown that the trajectory of the system can point toward and eventually reach the consensus state with $A$ or $B$, depending on whether $p_A \left( t^{\ast}\right) >  p_B \left( t^{\ast}\right)$ or $p_A \left( t^{\ast}\right) < p_B \left( t^{\ast}\right)$, where $t^{\ast} > 0$ is the critical time at which the representative point leaves the initial position.

\begin{figure}
\includegraphics*[width=17cm]{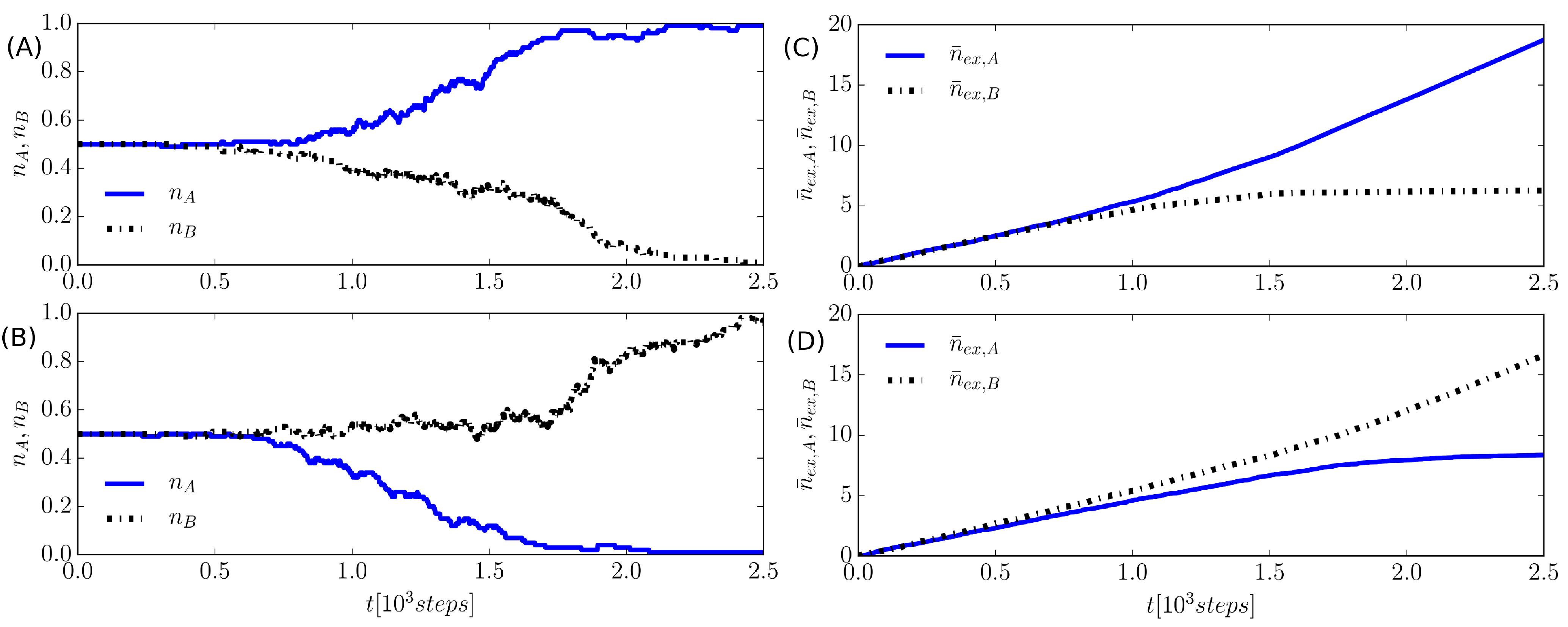}
\caption{
Results from two single simulations of a system with $N = 100$ agents, starting from SIC at $(x_0,y_0) = (0.5,0.5)$ and reaching two different consensus states about name $A$ or $B$.
Panels (A) and (B) show the population fractions  $x(t) = n_A(t)$ and $y(t) = n_B(t)$.
Panels (C) and (D) show the corresponding average number of examples recorded by an agent, $\bar{n}_{ex,A}(t)$ and $\bar{n}_{ex,B}(t)$.}
\label{fig:singleRuns}
\end{figure}

The convention $A$ or $B$ is selected randomly, depending on various factors related to the specific realization of the system evolution, such as the numbers of examples $\bar{n}_{ex,A}(t)$ and $\bar{n}_{ex,B}(t)$  recorded by the agents until time $t$, their quality from the point of view of the generalization,  and the initial asymmetry of the thresholds for generalizing, $n^{\ast}_{ex, A} \ne n^{\ast}_{ex, B}$. 
The asymmetrical thresholds $n^{\ast}_{ex, A} = 4 < n^{\ast}_{ex, B} = 5$ produce a bias toward consensus in $A$ and play  a crucial role in the subsequent Bayesian semiotic dynamics; in fact, swapping the threshold  values (setting $n^{\ast}_{ex,A} =5 > n^{\ast}_{ex,B}= 4$), the approach to consensus occurs with the outcomes $A$, $B$ swapped.

We observed that for $N \gtrsim N^{\ast} \approx 500$, the chances that the system converges to $(B)$ become negligible.
This can be seen in panels (C) and (D) of Fig. \ref{fig:singleRuns}, showing $\bar{n}_{ex,A}(t)$ and $\bar{n}_{ex,B}(t)$ \textit{versus} time (averaged over the agents of the system) for single runs, a population of $N = 100$ agents, and SIC, for different runs that relax toward consensus $A$ and $B$, respectively.
After an initial transient, in which $\bar{n}_{ex,A}(t) \approx \bar{n}_{ex,B}(t)$, they differ more and more from each other at times  $t > t^{\ast}$.
In turn, also $p_A$ and $p_B$ begin to differ significantly from each other, thus affecting the rate of depletion of the populations during the subsequent dynamics. 
For instance, if $p_A > p_B$, then $p_{B \rightarrow AB} > p_{A \rightarrow AB}$, see Eqs.~\eqref{eq:learningP},  which means that the depletion of $n_B$ occurs faster then that of $n_A$. In turn, this favours the decay of the mixed states $(A,B)$ into the state $(A)$, see Eqs.~\eqref{eq:agreementP}, being $n_A > n_B$.

\begin{figure}
\includegraphics*[scale=0.5]{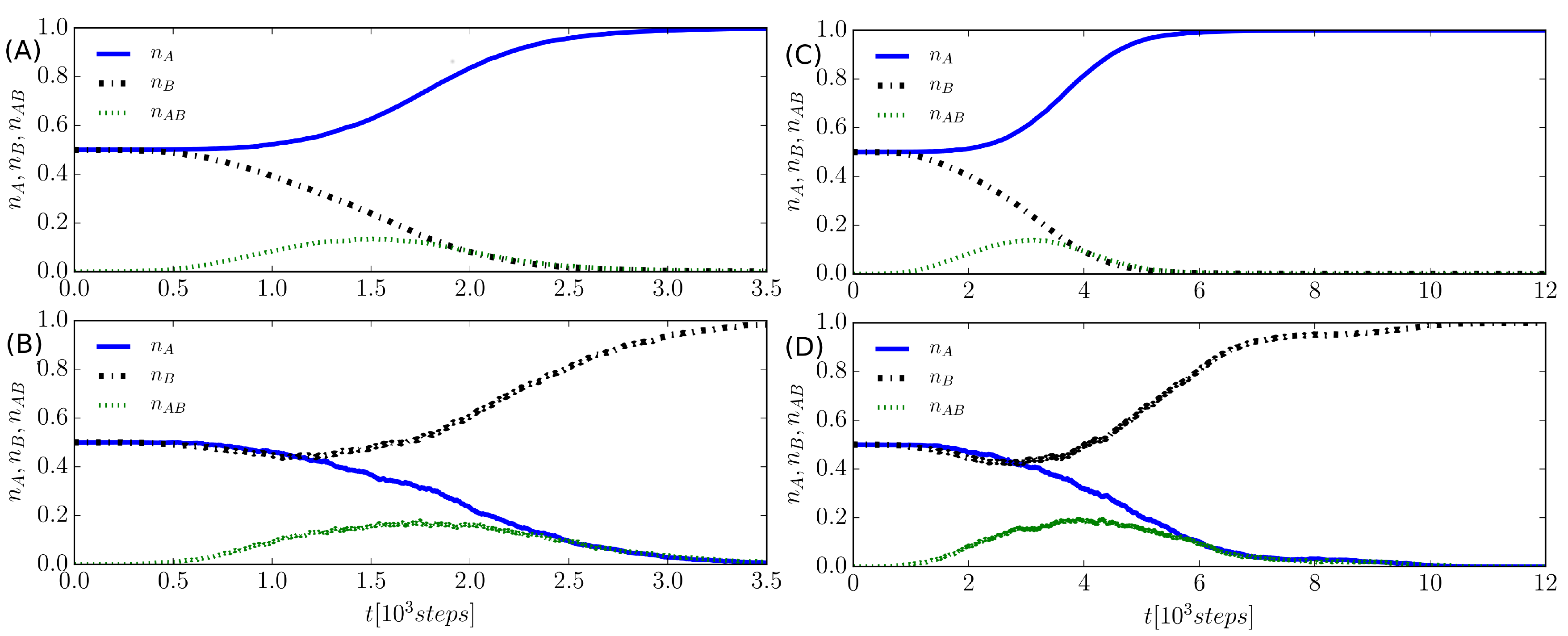}
\caption{
Population fractions $n_A(t)$, $n_B(t)$, and $n_{AB}(t)$, \textit{versus} time, starting from SIC; results are obtained by averaging over $600$ runs. 
Left column (panels (A) and (B)): a system with $N=100$ agents can reach consensus with name $A$ (panel(A), about $91\%$ of runs) or name $B$ (panel (B), about $9\%$ of runs).
Right column (panels (C) and (D)): system with $N=200$ agents, reaching consensus with name $A$ in about $96\%$ of runs (panel (C))) and with name $B$ in about in the remainder 4\% of runs (panel (D)).
} 
\label{fig:multipleruns1}
\end{figure}

The asymmetry discussed above also affects the convergence times $t^{A}_{conv}$ and $t^{B}_{conv}$ and we find $t^{B}_{conv} > t^{A}_{conv}$ in all the numerical simulations.
Despite the noise, such a trend is already appreciable in a single run, as shown in panels (A) and (B) of Fig. \ref{fig:singleRuns}.
The mean fractions $n_A(t)$, $n_B(t)$, and $n_{AB}(t)$ \textit{versus} time, obtained by averaging over many runs, result in less noisy outputs and provide a more clear picture of the difference, which is visible in Fig.~\ref{fig:multipleruns1}, obtained using $600$ runs starting with SIC and for $N = 100$ agents (panels (A) and (B)) and $N = 200$ agents (panels (C) and (D)). 
In addition, the convergence times depend on the system size, increasing with the number of agents $N$: compare the left panels (A) and (B), where $N = 100$ agents, with the right panels (C) and (D), where $N = 200$ agents.

The possibility that the same system, starting with the same initial conditions and evolving with the same dynamical parameters, can reach either $A$ or $B$ is a consequence of the stochastic nature of dynamics. 
This does not happen for $N \gtrsim N^{\ast}$, when both $\bar{n}_{ex,A}$ and $\bar{n}_{ex,B}$ reach some threshold values close to those observed at $t_{conv}$, which is clearly a value sufficient for the agents to generalizing concept $C$. 
In fact, the scaling law of $t_{conv}$ with $N$ shows that the sum of $\bar{n}_{ex,A}$ with $\bar{n}_{ex,B}$ becomes nearly constant for $N \gtrsim N^{\ast}$, implying that the dynamics is uniquely determined, that is, the consensus always occurs at $A$ from SIC, once the agents have stored a threshold number of $\bar{n}_{ex,A}$, $\bar{n}_{ex,B}$. 
It is found that these threshold values correspond to $\bar{n}^{\ast}_{ex,A}=21 $, $\bar{n}^{\ast}_{ex,B}= 12$.
Note that in $\bar{n}^{\ast}_{ex,A} $, $\bar{n}^{\ast}_{ex,B}$  we add values the four initial given examples stored in the agents' inventories at the beginning. 
The reason is that the generalization function $p(t)$ outputs will effectively depend on them all. 
Therefore, at these threshold values, it would be very unlikely that $p_B > p_A$, and so it would be the same for the consensus at $B$.

\begin{figure}
\includegraphics*[scale=0.5]{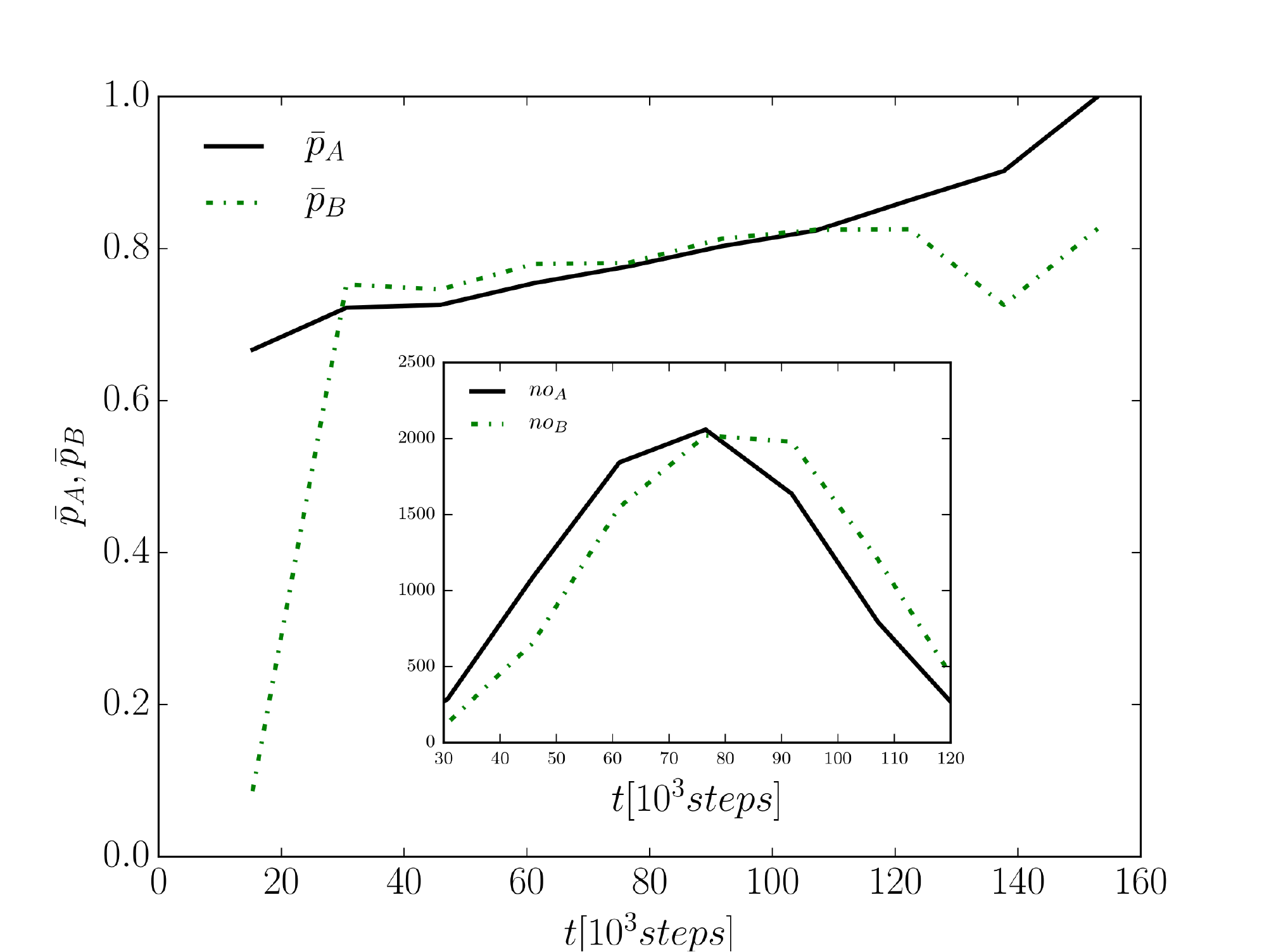}
\caption{Average values $\bar{p}_{A}(t)$ and $\bar{p}_{B}(t)$ computed using a (temporal) bin $\Delta t = 16  \times 10^{3}$, \textit{versus} time from for a single run of a system reaching consensus at $A$. 
The convergence time is $t_{conv}\approx 160 \times 10^{3}$ and the population size is $N=5000$. 
The inset shows the average number of learning attempts $no_A$, $no_B$  \textit{versus} time for the same single run. }
\label{fig:binned}
\end{figure}

Now we consider the Bayesian probabilities $p_A(t)$ and $p_B(t)$ computed by agents and the corresponding number of learning attempts $no_A(t)$ and $no_B(t)$ made by agents at time $t$ to learn concept $C$ in association with word $A$ or $B$, respectively, i.e. the number of times that the agents compute $p_A$ or $p_B$ (only the case of a system starting with SIC is considered).
We consider a single run of a system with $N=5000$ agents and study the average values $\bar{p}_{A}(t), \bar{p}_{B}(t)$, obtained by averaging $p_A(t)$ and $p_B(t)$ over the agents of the system.
We also assume a coarse-grained view, consisting in an additional average of $\bar{p}_{A}(t)$, $\bar{p}_{B}(t)$, and $no_A$, $no_B$, over a a temporal bin $\Delta t = 16 \times 10^{3}$, in order to reduce random fluctuations.
Figure \ref{fig:binned} shows the time evolution of the average probabilities $\bar{p}_{A}(t)$ and $\bar{p}_{B}(t)$ in the time-range where data allow a good statistics.
The probabilities grow monotonically and eventually reach the value one.
While this points at an equivalence between the mean-field regime of the Bayesian naming game and that of the two-conventions NG model, in which agents learn at the first attempt (one-shoot learning), such an equivalence is suggested but not fully reproduced by the coarse-grained analysis.
The time evolution of the number of learning attempts $no_A(t)$ and $no_B(t)$ shows that they are negligible both at the beginning and at the end of the dynamics --- see inset in Fig. \ref{fig:binned}.
This is due to the fact that at the beginning it is most likely that either interactions between agents with the same conventions take place (starting with SIC, each agent has a probability of 50\% to interact with an agent having the same convention) or interactions between agents with different conventions but with still too small inventories to be able to generalize concept $C$, leading to reinforcement processes only.
When approaching consensus, agents with one of the conventions constitute the large majority of the population and thus they are again most likely to interact through reinforcements only. 
Thus, the largest numbers of attempt to learn concept $C$ in association with $A$ and $B$ are expected to occur at the intermediate stage of the dynamics.
In fact,  $no_A(t)$ and $no_B(t)$ are observed to reach a maximum at $t \approx t_{conv}/2$ for any given system size $N$, as visible in the inset of Fig. ~\ref{fig:binned}. 
Notice that also the fraction of agents $n_{AB}$, who know both conventions and can communicate using both name $A$ and name $B$, possibly allowing other agents to generalize in association with name $A$ or $B$, reaches its maximum roughly at the same time.


\section{Conclusion}
\label{sec:conclusion}

We introduced a novel agent-based model that describes the appearance of linguistic consensus through a word-learning process.
The work presented is exploratory in nature, concerning the minimal problem of a single concept that can be associated to two different possible names $A$ or $B$, but is aimed at providing a prototype of general framework for describing the interaction between the social and the cognitive dimension.
To this aim, the model is constructed on the basis of the semiotic dynamics of the NG model and is then extended by adding a Bayesian cognitive process,  mimicking human learning processes. 

The model describes in a natural way (1) the uncertainty accompanying the first phase of a learning process, (2) the gradual reduction of the uncertainty as more and more examples are provided, and (3) the ability to learn from a few examples.
The semiotic dynamics of the synonyms is different from the basic NG, in that it depends on parameters that are of a strictly cognitive nature, such as the thresholds $n_{ex}^\ast$ of the number of examples necessary before an agent can try to generalize and the acceptance threshold $p^\ast$ for carrying out the generalization of a concept.
The interplay between the asymmetry of the conventions $A$ and $B$, the system size, and the stochastic character of the time evolution, have dramatic consequences on the consensus dynamics:
there is a critical time $t^\ast > 0$, when the system begins to move in the phase-plane to eventually converge toward a consensus state;
there is a critical system size $N^*$, such that for $N < N^\ast$ the system can end up in any of the two consensus states and the convergence times depends on $N$;
there is an asymmetry in the branching probabilities that the system converges toward one of the two possible conventions and of the corresponding convergence times; 
the scaling laws of the convergence times \textit{versus} $N$ differ from those observed in the basic NG model, because they depend on the learning experience of the agents. 

The cognitive dimension offers additional possibilities for modelling in terms of specific cognitive parameters problems that are out of the reach of traditional social dynamics models.
The model illustrated in this work represents a step toward a generalized Bayesian approach to social interactions, leading to cultural conventions.

Future work can address specific problems of current interest from the point of view of cognitive processes; or features relevant from the general standpoint of complexity theory.
In the first case, it is possible to study in the cognitive dimension the semiotic dynamics of homonyms, synonyms, and innovation, e.g., the cognitive conditions leading to a name $A_1$, associated to a concept $C_1$, splitting into two names $A_1$ and $A_2$, associated to two related but distinct concepts $C_1$ and $C_2$, as more examples become available that make the two concepts eventually distinguishable from each other --- a type of problems that cannot be tackled within models of cultural competition.
In the second case, one can mention the classical problem of the interplay between a central information source (bias) and the local influences of individuals --- this time in a cognitive framework.

Another question to be investigated within a cognitive framework would be the role of heterogeneity.
In fact, heterogeneity is known to characterize most of the known complex systems at various levels --- here the diversity could affect the dynamical parameters of e.g. the different competing names as well as those of the agents.
Heterogeneity of individuals can lead to counter-intuitive effects, such as resonant behaviors \cite{Tessone2009a,VazMartins2009a}.
Furthermore, the complex, heterogeneous nature of a local underlying social network can change drastically the co-evolution and the time-scales of the conventions in competition with each other \cite{Toivonen2009a}.

\section*{Acknowledgements}
The authors acknowledge support from the Estonian Ministry of Education and Research through Institutional Research Funding IUT (IUT39-1),
the Estonian Research Council through Grant PUT (PUT1356),
and the ERDF (European Development Research Fund) CoE (Center of Excellence) program through Grant TK133.

We also thank Andrea Baronchelli for providing useful remarks about the naming game model and the manuscript.

\end{document}